\documentstyle[12pt]{article}
\newlength{\natwide}
\newlength{\niwide}
\begin{document}
\setcounter{totalnumber}{99}
\renewcommand{\textfraction}{0}
\title{Coulombically Interacting Electrons in a One--dimensional
Quantum Dot}
\author{Wolfgang H\"ausler and Bernhard Kramer\\
Physikalisch--Technische Bundesanstalt Braunschweig\\
Bundesallee 100, 3300 Braunschweig F.~R.~G.}
\date{Received~:\rule{3cm}{0cm}}
\maketitle
\begin{abstract}
The spectral properties of up to {\em four} interacting electrons
confined within a quasi one--dimensional system of finite length
are determined by numerical diagonalization including the spin
degree of freedom. The ground state energy is investigated as a
function of the electron number and of the system length. The
limitations of a description in terms of a capacitance are demonstrated.
The energetically lowest lying excitations are physically explained as
vibrational and tunneling modes. The limits of a dilute,
Wigner--type arrangement of the electrons, and a dense, more
homogeneous charge distribution are discussed.
\end{abstract}
\mbox{}\\ \hspace*{1cm}{\it Introduction}\\[1mm]
Interaction effects play a crucial role in the understanding of
the electrical transport properties of very small condensed
matter systems at low temperatures \cite{kramer}. Examples are
\begin{itemize}
\item the Coulomb Blockade \cite{coulomb,zpb}, where charging energies of
single electrons suppress the current through a dissipatively shunted
tunnel junction
\item Single Electron Tunneling oscillations \cite{set} of the voltage
across such a junction at constant current
\item resonance--like oscillations of the conductance of quantum dots,
being periodic in multiples of the elementary charge inside the dot
\cite{meirav,kouwenhoven}
\end{itemize}
An important feature in all conductance measurements on tunnel
junctions and quantum dots is the relative isolation of the
sample region from the `external world'. In the blockade
experiments this is achieved by a shunt impedance representing
the (phase randomizing) influence of a coupling to the
electromagnetic environment. For quantum dots weak coupling is
achieved by highly resistive tunnel junctions. Here the time
scale for a change of the electron number is large compared to
the other inverse energies involved, namely the Fermi energy of
the external wires, the charging energy and the characteristic
energy of the dissipative heat bath. Therefore the electron
number on the time scale of all relaxation processes
approximately becomes a good quantum number. Due to its relative
isolation, the quantum properties of the disconnected dot can be
considered as one of the dominating factors in the single
electron phenomena. The Coulomb interaction should be taken into
account, because the charging energy is the most relevant energy
scale of the problem. The commonly used overall description by
means of a capacitance $\:C\:$ \cite{schoen,beenakker} is not
completely obvious \cite{eckernetal,corr91} and needs to be
justified. This is of particular relevance, if the number of
charge carriers is reduced to only about 10 in semiconductor
samples \cite{kouwenhoven,heitmann92}. Then both Coulomb
interaction and kinetic (confinement) energy start to play a
role. To clarify the interplay between these two energy contributions
\cite{johnson92} has been one of the main motivations for our
quantitative study of a model of $\:N\le 4\:$ interacting
electrons in a 1D square well potential. We shall see, that the
non--commutativity of the two corresponding energy operators has
an important influence on the ground state energy and on the
excitation spectrum. The ground state energy in general cannot be
obtained by the charging formula $\:Q^2/2C\:$ where $\:Q\:$ is
the total charge. The excitation energies have no similarity to
the discrete level structure pertaining to the confinement. We
shall present a qualitative picture, which allows to understand
physically the spectrum in the correlated case.

In the following Section describes our model and the
calculational method. The third Section addresses the question of
a capacitance definition in such correlated systems. One
possibility to investigate this question is to compare the
quantum mechanical ground state energy of $\:N\:$ interacting,
confined electrons with the charging energy of a corresponding
capacitor. In contrast to previous work \cite{bryant,inter,pfannkuche92}
we have been able to treat up to $\:N=4\:$ electrons with high
accuracy.

The excitation spectrum is discussed in the fourth
Section, for various electron densities.
An inhomogeneous distribution is established for low charge
density with essentially regularly
spaced elementary charges as a consequence of the dominating
Coulomb repulsion in that limit. We designate this arrangement of
electrons as a {\em Wigner molecule\/}, in analogy to the Wigner
electron lattice in dilute infinite systems with Coulomb
interaction. Two kinds of elementary excitations are identified,
{\em vibrational} and {\em tunneling} modes, that are
characteristic for dilute and intermediate densities
respectively.

According to experimental situations we conclude that for the
system parameters that characterize inversion layer based quantum
dots the quantum mechanics of the electron--electron interaction
cannot be completely discarded. In particular it should be
possible to observe the features related to the existence of the
low lying correlated excitations by suitable sample fabrication.

\mbox{}\\ \hspace*{1cm}{\it Interacting electrons in a square well
potential}\\[1mm]
For the electron--electron interaction, we use the form
\begin{equation}\label{lowdin}
V(x,x')\propto\frac{1}{\sqrt{(x-x')^2+\lambda^2}}\quad .
\end{equation}
which behaves Coulombically at large distances. $\:\lambda\:$ is
a measure for the width of the electron wave functions in
transversal direction. In most of our calculations we take $\:{\lambda
/L=2\cdot 10^{-4}\ll 1}\:$, where $\:L\:$ is the system length.
Then, the eigenvalues of the Hamiltonian
\begin{equation}\label{model}
H=E_{\rm H}\frac{a_{\rm B}}{L}\left( \frac{a_{\rm B}}{L}H_0+H_{\rm I}
\right)
\end{equation}
depend only weakly on $\:\lambda\:$. $\:E_{\rm H}=e^2/a_{\rm B}\:$ is
the Hartree energy, $\:a_{\rm B}=\varepsilon\hbar^2/me^2\:$ the Bohr
radius, $\:\varepsilon\:$ the relative dielectric constant and $\:m\:$
the electron mass. In most cases we present eigenvalues of the operator
enclosed by round brackets. The relative importance of the kinetic energy
in the 1D square well potential
\begin{equation}\label{h0}
H_0=\sum_{n,\sigma}\epsilon_nc^{\dagger}_{n,\sigma}
c^{\phantom{\dagger}}_{n,\sigma}
\end{equation}
($\:\epsilon_n\propto n^{2}\:$, $\:n\in\mbox{$\hspace{0.4\natwide}
\mbox{\sf N}\hspace{-1.25\natwide}\mbox{\sf I}\hspace{-\niwide}
\hspace{1.15\natwide}$}\:$) decreases as compared to the Coulomb energy
\begin{equation}\label{hi}
H_{\rm I}=\sum _{n_{1}\ldots n_{4},\sigma_{1},\sigma_{2}}
V_{n_{4}n_{3}n_{2}n_{1}}c^{\dagger}_{n_{4}\sigma_{1}}
c^{\dagger}_{n_{3}\sigma_{2}}
c^{\phantom{\dagger}}_{n_{2}\sigma_{2}}
c^{\phantom{\dagger}}_{n_{1}\sigma_{1}}
\end{equation}
with increasing system length $\:L\:$.
The matrix elements $\:V_{n_{4}n_{3}n_{2}n_{1}}\:$ are real and
do not depend on the electron spin $\:\sigma\:$. The total spin
$\:S\:$ is therefore conserved and all eigenvalues are
$(2S+1)$ -- fold degenerate. The symmetry relations are
$\:V_{n_{4}n_{3}n_{2}n_{1}}=V_{n_{4}n_{2}n_{3}n_{1}}=
V_{n_{1}n_{3}n_{2}n_{4}}=V_{n_{3}n_{4}n_{1}n_{2}}\:$ and
$\:V_{n_{4}n_{3}n_{2}n_{1}}=0\:$ if $\:\sum_in_i=\;\mbox{odd}\:$.
For $\:\lambda /L \ll 1\:$
\[
V_{n_{4}n_{3}n_{2}n_{1}}\approx
-8\pi\int_{0}^{\infty}\mbox{d}q\;\left (\ln
(q\lambda /L) + C\right )\hat{f}_{14}(q)\hat{f}_{23}(q)
\]
where $\:\hat{f}_{ij}\:$ is the Fourier transform of the product
$\:\varphi^{*}_{n_{i}}(x)\varphi^{\phantom{*}}_{n_{j}}(x)\:$ of
the one--electron eigenfunctions of $\:H_{0}\:$, and
$\:C=0.577\:$ the Euler constant.

For the numerical diagonalization single particle
states $\:c^{\dagger}_{n\sigma}\hspace{0.5ex}\rule[-0.5ex]{0.1ex}{0.9em}
\hspace{0.5ex}0\raisebox{-0.3ex}{\mbox{\large\tt >}}\:$ with $\:1\le n\le
M\:$ were chosen (usually $\:M=9\ldots 17\:$, depending on the
calculation). The properly anti--symmetrized, non--interacting
$\:N$--particle basis $\:\psi^{(N)}_{\nu}\:$, including the spin
degree of freedom, is of dimensionality
$\:\displaystyle R=\rule[-3ex]{0ex}{7.2ex}\left( 2M\atop N\right)\:$,
$\:1\le\nu\le R\:$.
In our calculations $\:R\:$ was restricted to $\:1.5\cdot 10^4\:$,
even when using Lanczos procedures. To avoid loops over all $\:R^2\:$
matrix elements of the Hamiltonian, we used the following economic
procedure to occupy the matrix. Starting from the $\:(N-2)$ -- particle
basis, the application of two creation operators onto a certain
$\:\psi^{(N-2)}_{\mu}\:$ generates say the $\:N$--particle state
$\:\psi^{(N)}_{\nu}\:$ with proper sign. $\:\psi^{(N)}_{\nu}\:$ corresponds
to a certain row $\:\nu\:$ of the Hamiltonian matrix. Creating from
the same $\:\psi^{(N-2)}_{\mu}\:$ a (different or the same~!)
$\:\psi^{(N)}_{\nu '}\:$ identifies a certain column $\:\nu '\:$.
The {\em independent} summation over all possible two--particle
excitations and subsequent summation over all $\:(N-2)$ -- particle
states generates eventually all non--vanishing entries (including
sums from $\:n_4=n_1\:$ and/or $\:n_3=n_2\:$,
cf.\ (\ref{hi})) of the Hamiltonian matrix.

Typical examples of $\:N$--electron energy spectra are shown in
Figure~\ref{fspect}. In presence of interaction, $\:N>1\:$, the density
of states is much more inhomogeneous as a function of the energy.
The lowest eigenvalues
form {\em multiplets} of extremely small width when $\:L\gg Na_{\rm B}\:$.
The total number of states within each of these
multiplets, including degeneracies, is $\:2^N\:$.

\mbox{}\\ \hspace*{1cm}{\it Ground state energies}\\[1mm]
Figure~\ref{e0n} shows the dependence of the ground state energy per
particle $\:E_{0}/N\:$ on the particle number $\:N\:$
for different $\:L\:$. The data are multiplied by $\:L\:$ in
order to eliminate the trivial $\:L$--dependence. The charging
model would yield a straight line for the ground state energy as a
function of the particle number $\:E_{0}(N)\:$ when plotted in
the same way. In very small systems $\:E_0/N\:$ deviates from a linear
$\:(N-1)$ -- dependence due to the discreteness of the spectrum of
$\:H_0\:$. On the other hand, for systems with large $\:L\:$ the
formation of an inhomogeneous charge density (Wigner molecule,
see below) prevents the ground state of few electrons energy to obey
$\:E_0/N\propto (N-1)\:$. A better approximation is obtained by
considering the Coulomb energy of $\:N\:$ point charges at equal
distances $\:r_{\rm s}=L/(N-1)\:$.

The importance of this charge ``crystallization'', which is
a consequence of the charge quantization, for the capacity per unit
length $\:C/L\:$ can be visualized for equidistant
point charges $\:e\:$ in 1D with a charge density distribution
\[
\varrho(x)=\sum_{j=1}^N\delta (x-x_j)\quad ,\quad x_j=\frac{j-1}{N-1}L
\quad .
\]
For a total charge $\:Ne\:$ the capacity of such an arrangement
is defined as
\[
C(N):=(Ne)^2/2U
\]
where
\[
U=e^2\sum_{i,j\atop i\ne j}\frac{1}{|x_i-x_j|}
\]
is the charging energy. For $\:N\ge 3\:$
\[
U=\frac{e^2}{L}(N-1)\sum_{j=1}^{N-1}\frac{j}{N-j}=
\frac{e^2}{L}N(N-1)\sum_{j=2}^{N}\frac{1}{j}
\]
and therefore
\[
C(N)/L=\frac{N}{2(N-1)}\left[ \sum_{j=2}^{N}\frac{1}{j}\right]^{-1}\quad .
\]
In contrast to the classical capacitance of a homogeneously
charged and long cylinder this capacitance per unit length is
independent on $\:L\:$ but explicitely dependent on the charge.
In particular for small $\:N\:$ the classical concept of a
capacitance obviously is inapplicable. Also in higher
dimensionalities we expect considerable fluctuations of the capacitance
as a function of the charge due to an inhomogeneous charge density.

On the other hand for short systems with a more homogeneous
charge distribution, quantum mechanical corrections to the ground
state energy also do not allow the use of the capacitance
formula. In Figure~\ref{e0l} $\:E_{0}/N\:$ is shown as a function
of $\:r_{\rm s}\:$. The deviations from the capacity like $\:1/L\:$
behavior occurring below $r_{\rm s}\mbox{\raisebox{-4pt}{$\,
\buildrel<\over\sim\,$}}50a_{\rm B}$ cannot be
attributed to a simple additive influence of the kinetic part of
the Hamiltonian. The ground state energy, of the non--interacting
system has already been subtracted in Figure~\ref{e0l}.

\mbox{}\\ \hspace*{1cm}{\it Low lying excitations}\\[1mm]
For $\:L\gg Na_{\rm B}\:$ the spectrum of the low lying excitations
can be understood using the picture of a Wigner molecule. The one
particle density shows $\:N\:$ approximately equidistant peaks
\cite{jaure}. One type of excitation in such an arrangement is of
a phonon kind due to the Coulomb forces between the charges.
Similar to the one particle density these excitations are
insensitive to the total spin and the symmetry properties of the
wave function.

To estimate the asymptotic behaviour of typical phonon
frequencies $\:\Omega\:$ as a function of the electron distance
$\:r_{\rm s}\:$, we assume $\delta$--functions or Gaussians for
in the charge density of each peak. The Gaussians can be
considered to emerge from the harmonic oscillator ground state
wave functions due to the linearized electrostatic potential for
one certain electron, leaving the other electrons fixed. The
result of this crude estimate is a $\:r_{\rm s}^{-\gamma}\:$
decrease of $\:\Omega\:$ with $\:\gamma =3/2\:$
($\delta$--function density) or $\:\gamma =1\:$ (Gaussian density).
Figure~\ref{oml} shows $\:\Omega L/a_{\rm B}\:$ as function of
$\:r_{\rm s}\:$ for different $\:N\:$, $\:\Omega\:$ is the distance
between the lowest two multiplets in our spectra. The behavior at
large $\:r_{\rm s}\:$ is consistent with $\:\gamma\mbox{\raisebox{-4pt}
{$\,\buildrel>\over\sim\,$}}1\:$.
This indicates that the charge density distribution is more
localized than a Gaussian. For $\:r_{\rm s}\mbox{\raisebox{-4pt}
{$\,\buildrel<\over\sim\,$}}100a_{\rm B}\:$
the strong deviations from the asymptotic behavior signalize the
breakdown of the Wigner molecule.

The multiplets consist of a series of energy levels which,
compared to $\:\Omega\:$, are extremely close to each other.
Each level is $(2S+1)$ -- fold degenerate since all eigenstates
of $\:H\:$ are simultaneously eigenstates to $\:\hat{S}^2\:$. For
$\:N\ge 3\:$ the states are in general not products of a spatial
part and a spin part \cite{stevens}. The Pauli principle only
requires their transformation according to the totally
antisymmetric irreducible representation of the symmetric group
$\:S_N\:$.

In the following a qualitative understanding for the fine
structure spectrum within one multiplet is developed, starting
from the notion of the Wigner molecule. Quantitative results will
be published elsewhere \cite{pocket}. The total potential,
including the interaction (\ref{lowdin}), has $\:N!\:$
equivalent minima in the configuration space which is a
$N$--dimensional hypercube $\:L^N\:$. There are no further equivalent
minima, because $\:S_N\:$ is the {\em only} symmetry group of the
problem. For sufficiently low electron densities (the
corresponding $\:r_{\rm s}\:$ will be quantified by numerical
results below), the probability amplitude of the eigenfunctions
is well localized around the vicinity of the potential minima.
The barrier of the Coulomb interaction in the 1D case separates
adjacent minima.

Each of the probability amplitude peaks may be taken as one of
the basis functions of a finite dimensional ``pocket state
basis'' \cite{huller} $\:\{\hspace{0.5ex}\rule[-0.5ex]{0.1ex}{0.9em}
\hspace{0.5ex}j\raisebox{-0.3ex}{\mbox{\large\tt >}}\}\:$,
with $\:1\le j\le N!\:$.
Matrix elements of the Hamiltonian in this basis
$\:\raisebox{-0.3ex}{\mbox{\large\tt <}}j\hspace{0.5ex}
\rule[-0.5ex]{0.1ex}{0.9em}\hspace{0.5ex}H\hspace{0.5ex}
\rule[-0.5ex]{0.1ex}{0.9em}\hspace{0.5ex}j'\raisebox{-0.3ex}
{\mbox{\large\tt >}}\equiv H_{jj'}\:$ mutually connect different
permutations of $N$--electron states and behave like
tunneling integrals. The smallness of the $\:H_{jj'}\:$ is
crucially required to approximate the true interacting
eigenfunctions of $\:H\:$ by eigenvectors of the Hamiltonian in
this truncated basis. They are given by certain linear
combinations of the $\:\{\hspace{0.5ex}\rule[-0.5ex]{0.1ex}{0.9em}
\hspace{0.5ex}j\raisebox{-0.3ex}{\mbox{\large\tt >}}\}\:$.
Correspondingly, the true
energy eigenvalues are approximated by eigenvalues of the pocket
state Hamiltonian matrix. Since the amplitudes of the
$\:\hspace{0.5ex}\rule[-0.5ex]{0.1ex}{0.9em}
\hspace{0.5ex}j\raisebox{-0.3ex}{\mbox{\large\tt >}}\:$
decrease roughly exponentially with the distance
from its center, the $\:H_{jj'}\:$ have also this property and
thus the pocket state approach improves with increasing
$\:r_{\rm s}\:$. It is important to note, that this description is
not based on any single particle basis set as a starting point.
It becomes increasingly reliable with increasing influence of the
interaction. However, the diagonalization of the Hamiltonian
matrix in this basis cannot be achieved by Fourier
transformation, since $\:S_N\:$ is not abelian.

As a further approximation, we neglect all $\:H_{jj'}\:$ except
those which correspond to nearest neighbours in the configuration
space $\:L^N\:$. For sufficiently large $\:r_{\rm s}\:$ this is
justified by the exponentially fast decay of the pocket state far
away from its center. Due to the symmetry of the problem, all of
the remaining tunneling integrals $\:H_{jj'}=t\:$ must be equal
and are the only nontrivial entries in the Hamiltonian matrix.
Therefore the differences between the eigenvalues scale with the
common factor $\:t\:$.

In Figure \ref{ldelt} the energy difference $\:\Delta\:$ between
the two lowest numerically obtained eigenvalues is plotted versus
the system length $\:L\:$. The assumption of an asymptotically
exponential decay of the pocket states suggests that
$\:\Delta\propto \exp (-L/L_{\Delta})\:$ with
$\:L_{\Delta}\approx 1.5a_{\rm B}\:$ from Figure~\ref{ldelt}. This
establishes a length scale beyond which the spectrum of
non--interacting electrons is changed into one showing the multiplet
structure that is characteristic for the influence of the Coulomb
inter\-action. The tunneling energies $\:\Delta\:$ depend on
$\:\lambda\:$, because the height of the barriers between the
potential minima are proportional to $\:\lambda^{-1}\:$
(cf.~(\ref{lowdin})).

\mbox{}\\ \hspace*{1cm}{\it Conclusion}\\[1mm]
We have calculated numerically the energy spectra of up to
$\:N=4\:$ electrons confined in a quasi--one dimensional square
well potential of finite length. The discussion in terms of the
pocket state basis suggests that our classification of
the energy eigenvalues should remain valid also in 2D or 3D
systems, if the width of the system does not exceed the width of
one pocket state wave function. For $\:N=2\:$ we can reproduce the
fine structure in the lowest multiplet calculated by
Bryant \cite{bryant} with its sequence $\:S=0,1,0\:$ of total spins
and with equal level distances for sufficiently large size of a
rectangular two dimensional quantum dot, which is 10 times
longer than wide.

We have demonstrated that the ground state energies
$\:E_{0}(N)\:$ deviate from the $N^{2}$ -- behavior assumed in
the charging model because of the formation of a Wigner
molecule like structure at sufficiently low electron densities
(quantization of the charge) and the quantum mechanical
influence of the kinetic energy (non--commutativity of $\:H_0\:$
and $\:H_{\rm I}\:$). Only in sufficiently large systems {\em and}
at sufficiently high electron densities, a capacitance--like
behavior can be obtained.

We have obtained 3 different regimes for the electron densities
to characterize the excitation spectra. The Wigner molecule is
found to be fully established for densities $\:1/r_{\rm s}\:$ below
$\:10^{-2}a^{-1}_{\rm B}\:$ (Figure~\ref{oml}, see also
\cite{jaure}). Nevertheless, the description of the interacting
spectrum in terms of the pocket state picture does already hold
at much larger electron densities. Only for $\:L\mbox{\raisebox{-4pt}
{$\,\buildrel<\over\sim\,$}}
L_{\Delta}\approx 1.5a_{\rm B}\:$ the confinement energy starts to
dominate the Coulomb energy and the spectrum approaches the non--
interacting limit (Figure ~\ref{ldelt}). Neither the ground state
energy nor the level spectrum is given by a sum of kinetic and
potential energy eigenvalues separately~!

Experiments are frequently performed on AlGaAs/GaAs -- based
heterostructures which rather correspond to a 2D situation. It
is not obvious in how far our 1D classification for the length-- and
energy scales of few Coulombically interacting electrons can
be applied to that case. If we assume that at least the
qualitative aspects of our classification into different regimes
for the electron density can be used, the intermediate regime
should apply in most circumstances. Given the geometry and the
electron numbers in typical quantum dots \cite{meirav} (area of
the dot $\:\approx 10^5$nm$^2\:$, number of electrons
$\approx 10^2$ effective mass $\:\approx 0.07m\:$, dielectric
constant $\:\approx 10\:$) a mean distance of $\:r_{\rm s}\approx
3a_{B}\:$ can be estimated. For this relatively high electron
density and number the {\em ground state energy}, which is in first
approximation the relevant quantity that enters a dc--transport
experiment, can roughly be estimated by using the charging model.
However, the {\em excitation energies} are qualitatively
different from the ones expected within the non--interacting
picture. They are importantly characterized by the fine structure
level spectrum and the total spin. In experimental situations as
they have been realized recently by Meurer et.~al.\ \cite{heitmann92}
with only a few electrons per quantum dot, the charging model
even cannot be expected to yield correct results for the ground
state energy.

It should be possible to observe the excitation spectrum obtained
in this paper in optical measurements \cite{heitmann,heitmann92}
(if the potential is not strictly harmonic \cite{pfannkuche92})
and in non--linear transport measurements \cite{johnson92} at low
temperatures. At least for quasi 1D geometries, where the spatial
width of the dot region is of comparable size or smaller than the
width of the pocket state wave function, we predict a close
relation between the energies of low lying excitations and the
number of electrons in the dot. Only a finite number of tunneling
type excitations should exist for fixed electron numbers. The
energy scale for a 100 nm long 1D structure on AlGaAs/GaAs basis
can be estimated from Figure~\ref{fspect} to be $\:\Delta\approx
0.1$ --- $0.5\:$meV for $\:N=2\ldots 4\:$. However, this value
still depends on the effective width of the structure.

For the non--linear transport measurements two remarks should be
made. \ {\em i}) the excitation spectra of $\:N$-- and
$\:(N+1)\:$-- electron states are completely different, even the
typical splittings may differ by almost an order of magnitude.
They are not equidistant and independent on $\:N\:$. \ {\em ii})
the total spin of the electrons inside the dot region can be
changed only by $\:\pm 1/2\:$ when an electron is added or
removed at the finite conductance situation. This reduces the
number of transport channels available. For example the
conductance channel connected with the energy difference
$\:E^{(N=3)}_{S=3/2}-E^{(N=2)}_{S=0}\:$ ($\:\mu_d(N=2)\:$ in the
nomenclature of Ref.~\cite{johnson92}, $\:S\:$ is the total spin)
should not appear.\\[5mm]
{\em Acknowledgement :} We thank Kristian Jauregui for
enlightening discussions and constructive criticisms.
Furthermore we want to acknowledge the fruitful and exciting
conversations with all members of the Research Workshop on
Few--Electron Nanostructures in Noordwijk. This work was supported by
grants of the Deutsche Forschungsgemeinschaft (AP 47/1--1) and
EEC (Science Contract No.\ SCC$^*$--CT90--0020).
\newpage
\newcommand{\pap}[5]{#1, \ #2 {\bf #3}, #4 (19#5)}
\newcommand{\bk}[4]{#1~:\ {\it ``#2''}\\ #3 (19#4)}

\newpage
{\bf Figure Captions}\\[8mm]
\begin{figure}[h]
\parbox[b]{14cm}
{\caption[fspect]{\label{fspect}
Typical spectra of model (\ref{model}) for various $\:N\:$ and
$\:L=9.45a_{\rm B}\:$. For $\:N\ge 2\:$ the low lying
eigenvalues form groups of (fine structure) multiplets, the total
number of states per multiplet being equal to the dimensionality of
the spin Hilbert space $\:2^N\:$. For clarity the lowest multiplets
are magnified indicating the total spin of each level.
The ground state energy is subtracted respectively.
}}
\end{figure}
\begin{figure}[h]
\parbox[b]{14cm}
{\caption[e0n]{\label{e0n}
Ground state energies per particle $\:E_{0}/N\:$ multiplied by
$\:L/a_{\rm B}\:$ versus the particle number $\:N\:$
for $L=6.61a_{\rm B}$ ($\Box$) $L=16.1a_{\rm B}$ ($\circ$)
$L=94.5a_{\rm B}$ ($\triangle$) $L=944.8a_{\rm B}$ ($+$).
($\:\times\:$) denote the energy of $\:N\:$ fixed point
charges equally spaced at distances $\:L/(N-1)\:$.
The quantum mechanical ground state energies (slowly) approach
these values as $\:L\to\infty\:$.
}}
\end{figure}
\begin{figure}[h]
\parbox[b]{14cm}
{\caption[e0l]{\label{e0l}
Ground state energy per particle multiplied by $\:L/a_{\rm B}\:$ as a
function of the electron distance $\:{r_{\rm s}:=L/(N-1)}\:$;
the corresponding non--interacting ground state energy is subtracted.
Pronounced deviations from the Coulombic $\:1/L\:$ behavior occur below
$\:r_{\rm s}\mbox{\raisebox{-4pt}{$\,\buildrel<\over\sim\,$}}
50a_{\rm B}\:$. These deviations cannot be attributed
to a simple additive influence of the kinetic part of the Hamiltonian.
}}
\end{figure}\vspace*{\fill}\newpage
\begin{figure}[h]
\parbox[b]{14cm}
{\caption[oml]{\label{oml}
Energy difference $\:\Omega\:$ between the two lowest multiplets,
multiplied by $\:L/a_{\rm B}\:$ versus the mean particle distance
$\:r_{\rm s}\:$
for $\:N=2$ and $N=3$.
For $\:r_{\rm s}\mbox{\raisebox{-4pt}{$\,\buildrel>\over\sim\,$}}
100a_{\rm B}\:$ the asymptotic behaviour is recovered.
}}
\end{figure}
\begin{figure}[h]
\parbox[b]{14cm}
{\caption[ldelt]{\label{ldelt}
Logarithm of the energy difference $\:\Delta\:$ between the
ground state and the first excited state within the lowest
multiplet versus the system length for $N=2$, $M=11$ ($\Box$),
$N=3$, $M=13$ ($\circ$), and $N=4$, $M=10$ ($\triangle$). From
the slope we estimate $L_{\Delta}\approx 1.5a_{B}$.
}}
\end{figure}\vspace*{\fill}
\end{document}